\documentclass[sigconf,screen,nonacm]{acmart}

\usepackage{graphicx}
\usepackage{booktabs}
\usepackage{amsmath}
\usepackage{multirow}
\usepackage{xcolor}
\usepackage{subcaption}
\usepackage{placeins}
\usepackage{float}
\usepackage{makecell}
\usepackage{afterpage}
\usepackage{balance}

\newcommand{\method}{V-SPLADE}
\newcommand{\flops}{\textsc{FLOPs}}

\acmConference[Preprint]{Preprint}{2026}{}
\copyrightyear{2026}
\acmYear{2026}

\renewcommand\footnotetextcopyrightpermission[1]{}


\title{Inference-Free Multimodal Learned Sparse Retrieval \\ for Production-Scale Visual Document Search}

\author{Gyu-Hwung Cho}
\email{gyuhwung.cho@navercorp.com}
\affiliation{%
  \institution{NAVER Corp.}
  \city{Gyeonggi-do}
  \country{Republic of Korea}
}
\affiliation{%
  \institution{Seoul National University}
  \city{Seoul}
  \country{Republic of Korea}
}

\author{Youngjune Lee}
\email{yjlee511@gmail.com}
\affiliation{%
  \institution{NAVER Corp.}
  \city{Gyeonggi-do}
  \country{Republic of Korea}
}

\author{Kiyoon Jeong}
\email{kiyoon.jeong@navercorp.com}
\affiliation{%
  \institution{NAVER Corp.}
  \city{Gyeonggi-do}
  \country{Republic of Korea}
}

\author{Siyoung Lee}
\email{c.young@navercorp.com}
\affiliation{%
  \institution{NAVER Corp.}
  \city{Gyeonggi-do}
  \country{Republic of Korea}
}

\author{Sanggyu Han}
\email{sanggyu.han@navercorp.com}
\affiliation{%
  \institution{NAVER Corp.}
  \city{Gyeonggi-do}
  \country{Republic of Korea}
}

\author{Herv\'{e} Dejean}
\email{herve.dejean@naverlabs.com}
\affiliation{%
  \institution{Naver Labs Europe}
  \city{Meylan}
  \country{France}
}

\author{St\'{e}phane Clinchant}
\email{stephane.clinchant@naverlabs.com}
\affiliation{%
  \institution{Naver Labs Europe}
  \city{Meylan}
  \country{France}
}

\author{Seung-won Hwang}
\authornote{Corresponding author.}
\email{seungwonh@snu.ac.kr}
\affiliation{%
  \institution{Seoul National University}
  \city{Seoul}
  \country{Republic of Korea}
}

\renewcommand{\shortauthors}{Cho et al.}

\begin{document}

\begin{abstract}
As large-scale visual-document corpora such as arXiv papers and enterprise PDFs continue to grow, visual-document retrieval has gained increasing attention; yet it still lacks a deployable system that lexically indexes visual documents to serve queries without neural encoding at scale. Existing methods either achieve strong retrieval quality with VLM-based dense or multi-vector models but require neural query encoding at serving time, or avoid query encoding with OCR- or caption-based BM25 at the cost of time-consuming text extraction or generation. To fill this missing serving regime, we present \textbf{\method{}}, an inference-free sparse retriever for visual-document retrieval. However, such inference-free multimodal learned sparse retrieval systems remain underexplored and have not yet shown dense-level effectiveness under high sparsity. We attribute this limitation to a \textbf{lexical grounding problem}: visual sparse representations often fail to capture the lexical content embedded in document images. To address this problem, we introduce \textbf{caption-gated token supervision}, a training-only signal that uses VLM-generated captions as lexical cues to activate retrieval-relevant vocabulary dimensions. With this supervision, \method{} improves average NDCG@5 across six visual-document retrieval benchmarks by +13.8pp over the same-scale dense baseline and by up to +6.3pp over OCR- or caption-based BM25 baselines. On an 18.7M-document corpus, it more than doubles R@5 over the same-scale dense baseline and further improves competing retrievers through score fusion by up to +2.4pp R@5. Code will be released soon at \url{https://github.com/naver/v-splade}.
\end{abstract}

\maketitle
\section{Introduction}
\label{sec:introduction}

Large-scale visual-document corpora are becoming increasingly common~\cite{laurencon2024idefics3,arxiv2024annualreport}, but visual-document retrieval---the task of retrieving image-based documents for a user's text query---still lacks a production-scale lexical retriever~\cite{faysse2025colpali,shorten2026irpapers}.
The missing operating point is a retriever that serves text queries without neural query encoding while directly indexing visual documents.
Such a system would make large-scale visual-document retrieval more cost-effective.
However, current approaches satisfy only part of this requirement.
End-to-end dense and multi-vector VLM retrievers operate directly on visual documents and achieve strong accuracy, but require neural query encoding, larger backbones, or expensive multi-vector scoring at serving time~\cite{colbert,moreira2026nemotron,faysse2025colpali}.
Conversely, OCR- or caption-based BM25~\cite{robertson2009bm25} provides query-encoding-free lexical retrieval, but only after each visual document is converted into text through a time-consuming OCR pipeline or costly caption generation~\cite{faysse2025colpali,shorten2026irpapers}.

This leaves a missing operating point for large-scale visual-document retrieval: a lexical retriever that directly indexes visual-documents and serves queries without neural encoding.
Inference-free multimodal learned sparse retrieval (MMLSR) naturally fits this regime.
A VLM-based sparse encoder can map visual-documents directly into lexical sparse embeddings without OCR or caption generation, while Bag-of-Words (BoW) queries enable inverted-index retrieval without neural query encoding.
Yet MMLSR remains largely underexplored for visual-documents, especially in its inference-free form.
Although several notable efforts have developed MMLSR methods for general text-to-visual retrieval~\cite{nguyen2024expansion,song2025dense_sparse}, existing approaches have not yet matched the performance of comparable-scale dense models while preserving high sparsity.
This gap is more pronounced in the ViDoRe leaderboard, a widely used benchmark for visual-document retrieval systems, which contains no MMLSR entry as of writing~\cite{vidore_leaderboard}.

We attribute this limitation to a \textbf{lexical grounding problem} in sparse visual-document retrieval.
By lexical grounding, we mean the ability of a multimodal learned sparse retriever to map visually presented lexical evidence, in a rendered page image such as words and numbers, to the relevant lexical dimensions in its sparse representation.
In text sparse retrieval, this mapping is direct because document words are provided as input tokens, which naturally anchor vocabulary-indexed outputs.
In visual sparse retrieval, however, the same lexical content is observed only as pixels.
The encoder must therefore infer which vocabulary dimensions to activate without explicit text-token anchors.
When this grounding fails, the sparse representation may miss important document terms or activate spurious dimensions, making lexical matching less reliable.
We make this gap observable through a diagnostic study on rendered text documents.

\begin{figure}[t!]
\centering
\includegraphics[width=\columnwidth]{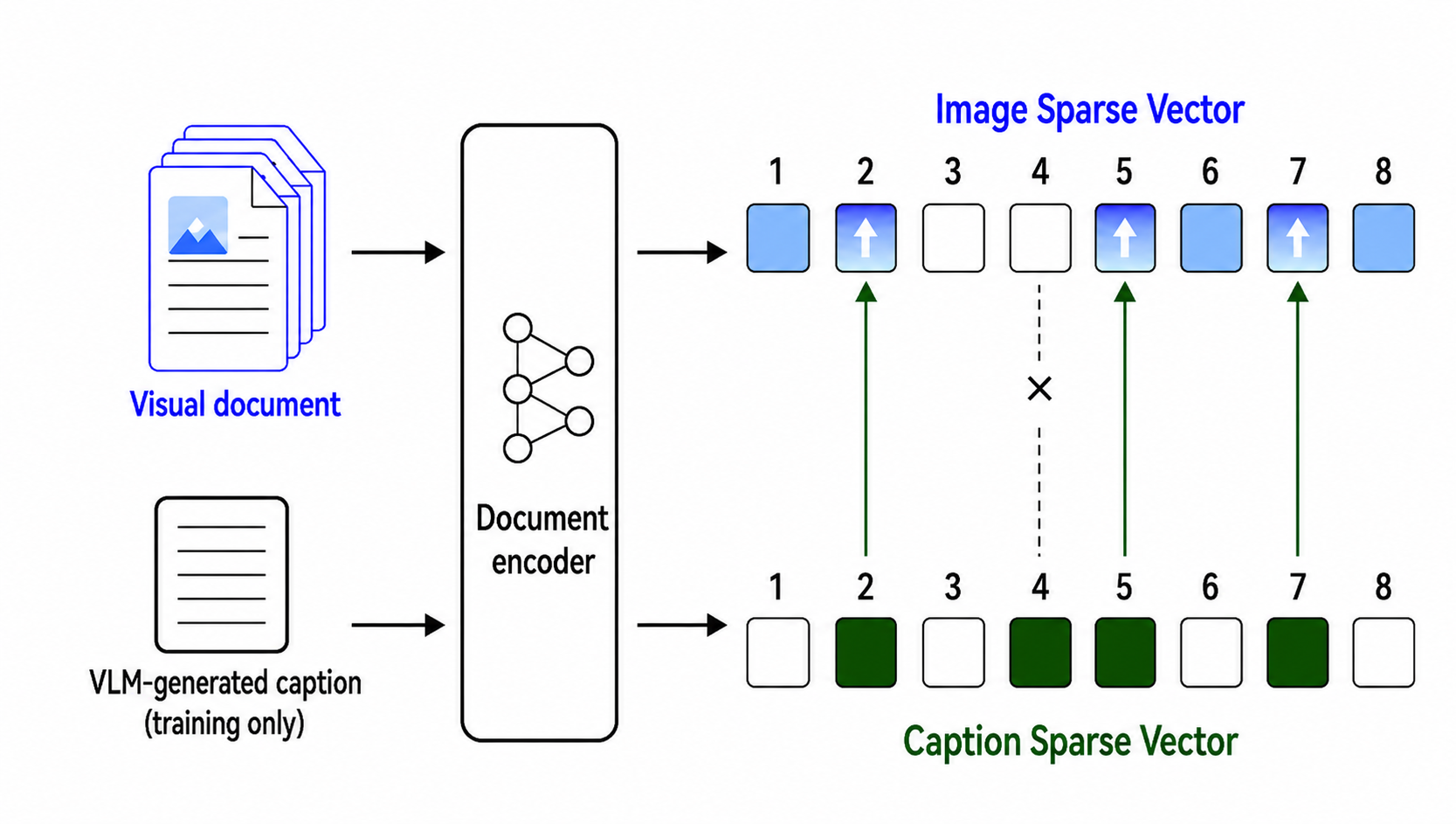}
\vspace{-8pt}
\caption{Caption-gated token supervision overview.}
\label{fig:overview}
\end{figure}

To address this problem, we introduce \textbf{caption-gated token supervision}, a training-only signal that uses VLM-generated captions to provide lexical token-level cues for visual sparse representations.
As shown in Figure~\ref{fig:overview}, the visual-document and its offline-generated caption are encoded into the same sparse vocabulary space.
The caption vector gates the image vector, reinforcing dimensions supported by both sparse views as reliable lexical evidence on the image side.
The caption branch is used only during training; at inference, the encoder maps each visual-document directly to an image-side sparse vector.

With this supervision, we introduce \textbf{\method{}}, a lexically grounded sparse retriever for the missing serving regime in visual-document retrieval.
\method{} maps each visual-document directly into a vocabulary-indexed sparse representation using a compact 250M visual-to-sparse encoder, without OCR or caption generation.
At serving time, sparse document vectors are stored in a standard inverted index, while query-token weights are served by learned token lookup without neural query encoding.
Furthermore, the lexical nature of these representations makes \method{} more robust under corpus scaling and complementary to dense retrievers, yielding gains both as a standalone retriever and as a fusion component.

\method{} consistently outperforms the main target-regime baselines in both standard benchmarks and large-scale retrieval.
Across six visual-document retrieval benchmarks, \method{} improves average NDCG@5 by +13.8pp over BiModernVBERT, a state-of-the-art compact dense retriever for visual-document retrieval in the same model-size regime, and by +5.7pp over the OCR-based lexical baseline.
On an 18.7M-document corpus, \method{} reaches R@5=0.228, compared with 0.090 for the same-backbone dense retriever.
It supports sub-10ms exact inverted-index search, approximate inverted-index search with latency comparable to HNSW~\cite{malkov2020hnsw}, and document encoding over $20\times$ faster than caption generation or OCR-based lexical pipelines.
As the corpus grows, \method{} retains recall more robustly than the dense baseline.
Through score fusion, it further improves competitive dense retrievers by up to +2.4pp R@5 on the 18.7M corpus.
Together, these results position \textbf{\method{} as a scalable lexical retrieval layer: fast to index, efficient to serve, robust under corpus scaling, and complementary to existing retrieval systems}.

In summary, we make the following contributions:
\begin{enumerate}
    \item We diagnose a lexical grounding problem in sparse visual-document retrieval.
    \item We propose caption-gated token supervision to lexically ground visual sparse representations.
    \item We develop \method{}, a sparse retriever for the missing serving regime in production-scale visual-document retrieval.
\end{enumerate}

\section{Related Work}
\label{sec:related_work}

\paragraph{Visual document retrieval.}
Visual-document retrieval has traditionally relied on OCR-extracted text for retrieval, but has increasingly shifted toward VLM retrievers that operate directly on rendered pages~\cite{faysse2025colpali}.
The strongest line has largely followed ColBERT-style late interaction, preserving fine-grained page evidence through multi-vector representations and MaxSim-style scoring~\cite{colbert,faysse2025colpali,moreira2026nemotron,gunther2025jinav4}.
While effective for complex pages, this regime is costly as a full-corpus first-stage retriever because it requires neural query encoding and scoring over hundreds to roughly a thousand visual tokens per document.
Recent compact backbones and reduced-token late-interaction models improve efficiency~\cite{teiletche2025modernvbert,masry2025colflor,xiao2025metaembed}, but still differ from the query-encoding-free lexical serving regime targeted by \method{}.

\paragraph{Text learned sparse retrieval.}
Text learned sparse retrieval provides the closest template for query-encoding-free lexical serving~\cite{formal2021splade,formal2021spladev2,gao2021coil,dai2020deepct,mallia2021deepimpact,lin2021unicoil}.
Among these methods, SPLADE~\cite{formal2021splade,formal2021spladev2} is the most prominent example: it maps text into sparse vectors over the language-model vocabulary, enabling weighted lexical matching with inverted indexes.
Because SPLADE representations live in vocabulary space, they can be paired with BoW-style queries to avoid neural query encoding.
Li-LSR~\cite{nardini2025lilsr} further strengthens this inference-free query side by replacing naive BoW weights with learned token-level lookup weights.
However, these successes have been largely confined to text, where input tokens directly anchor sparse vocabulary dimensions; rendered visual documents lack such anchors.

\paragraph{Multimodal learned sparse retrieval and caption supervision.}
Multimodal learned sparse retrieval has seen several notable research efforts, but remains far less established than text learned sparse retrieval~\cite{luo2023lexlip,chen2023stair}.
Notable MMLSR work has used dataset-provided captions for stable sparse expansion control~\cite{nguyen2024expansion} or sparse--dense inter-score self-distillation to improve multimodal sparse representations~\cite{song2025dense_sparse}.
However, these efforts have not yet achieved performance comparable to same-size dense models while maintaining high sparsity, and have rarely been applied to visual-document retrieval.
At the same time, work on generated textual descriptions suggests a promising source of lexical supervision.
Caption-based BM25 provides competitive lexical baselines for visual documents~\cite{nakata2024rethinking,shorten2026irpapers}, and generate-and-encode methods such as SERVAL show that VLM-generated document descriptions can act as strong semantic proxies~\cite{nguyen2025serval}.
MLLM-generated captions have also been used as auxiliary signals beyond visual documents, such as in text-video retrieval~\cite{lee2025caredpo}.
These results suggest that captions provide useful lexical and semantic evidence for retrieval.
This evidence motivates \textbf{\method{}} and its core training signal, \textbf{caption-gated token supervision}.

\section{Motivation: Lexical Grounding}
\label{sec:lexical-grounding}

This section examines a lexical grounding problem in sparse visual-document retrieval.
Text learned sparse retrieval is grounded in text by construction, because the input itself is a sequence of lexical tokens.
However, in sparse visual-document retrieval, the same lexical content appears only as visual evidence in a rendered page image.
A multimodal sparse encoder must therefore ground visual evidence into lexical vocabulary dimensions without explicit text-token anchors.
We refer to the challenge of activating vocabulary dimensions that correspond to lexical evidence observed in the page image as the \emph{lexical grounding problem}.

To make this problem observable, we conduct a controlled diagnostic experiment on 1,000 RLHN text documents~\cite{thakur2025rlhn}, a well-curated text IR dataset.
For each document, we create two aligned input views with the same underlying content. First, we keep the original source text as a text input. Second, we render the same text into a PDF-like page image and use the rendered page as a visual-document input.
We then pass each view through a vision-language backbone with an LM head, which projects the representation into language-model vocabulary space, and inspect the vocabulary dimensions with the highest activation values.
For a given top-$k$ set of activated dimensions, we measure the fraction of dimensions that overlap with the source-text BoW.
In summary, the diagnostic asks the following question: \emph{when the same document content is provided either as source text or as a rendered page image, how differently is its lexical content recovered in sparse vocabulary space?}

We use ModernVBERT as the diagnostic encoder because its MLM-based modality alignment trains it to recover text tokens from rendered document image inputs. This lets us examine how much of the lexical grounding problem remains under a model explicitly trained to recover lexical information from visual documents.

We compare four views:
\begin{enumerate}
    \item \textbf{Source text} as \emph{text}: the original source text, which serves as a text-side upper bound.
    \item \textbf{Rendered image} as \emph{image}: the PDF-like page image rendered from the same source text.
    \item \textbf{VLM-generated caption} as \emph{caption}: a caption of the rendered image, produced by Qwen3-VL-30B~\cite{qwenteam2025qwen3vl} using the same prompt as ColPali~\cite{faysse2025colpali} to extract visual-document information.
    \item \textbf{Image--caption gate} as \emph{gate}: the element-wise product between the image sparse representation from (2) and the caption sparse representation from (3).
\end{enumerate}

\begin{table}[h!]
\centering
\begin{tabular}{lccc}
\toprule
Representation & top-30 & top-50 & top-100 \\
\midrule
text (upper bound) & 0.974 & 0.858 & 0.518 \\
image & 0.560 & 0.445 & 0.301 \\
caption text & 0.574 & 0.509 & 0.355 \\
\textbf{gate (image $\times$ caption)} & \textbf{0.794} & \textbf{0.636} & \textbf{0.394} \\
\bottomrule
\end{tabular}
\caption{Diagnostic lexical grounding analysis: source-text BoW overlap among top-$k$ activated vocabulary dimensions.}
\label{tab:lexical-grounding}
\end{table}

Table~\ref{tab:lexical-grounding} shows a large gap between the source-text upper bound and the rendered-image representation. At top-30, the overlap drops from 0.974 for source text to 0.560 for the rendered image. This indicates that the lexical content remains visually present in the page, but is only partially recovered by the image-side sparse encoder. As a result, query matching must rely on a smaller and noisier set of lexical dimensions, which may explain why high-sparsity multimodal learned sparse retrieval is difficult to train.

Caption text partially improves lexical overlap, but remains insufficient on its own: at top-30, it reaches 0.574, slightly above the image representation at 0.560. Although prompt tuning optimized for this specific task may further improve the caption-only result, this setting also reflects a realistic condition where captions can be noisy or imperfectly aligned with the original document content.

In contrast, a product gate between the image sparse representation and the caption sparse representation yields a much stronger overlap, reaching 0.794 at top-30.
This suggests that the two views provide complementary evidence: the image representation carries meaningful but noisy lexical evidence from the rendered page while the caption representation provides a sharper lexical prior. This motivates caption-gated token supervision, which reinforces image-side vocabulary dimensions supported by the caption signal only during training.

\section{Method}
\label{sec:method}

\subsection{Overview}
\label{sec:method_overview}

Before describing how caption-gated token supervision addresses the lexical grounding problem, we first define the basic architecture of \method{}. This architecture combines SPLADE-based~\cite{formal2021spladev2} lexical sparse representations with Li-LSR-style~\cite{nardini2025lilsr} token lookup, forming an inference-free MMLSR retriever for the missing operating point of visual-document retrieval: direct visual-document indexing with query-encoding-free serving. We then introduce caption-gated token supervision as a training-only signal that lexically grounds image-side sparse activations produced by \method{}.

\subsection{\method{}}
\label{sec:architecture}
\label{sec:query_encoder}

\method{} extends SPLADE for direct visual-document indexing with a backbone choice tailored to the visual-document setting.
Prior MMLSR methods~\cite{nguyen2024expansion,song2025dense_sparse} often start from CLIP-style image--text encoders and learn a sparse output space on top of them. In contrast, \method{} uses a vision-language backbone with an LM head, which better matches the SPLADE formulation because its hidden states are already aligned with vocabulary prediction.

\paragraph{Image-side sparse representation.}
Following SPLADE~\cite{formal2021spladev2}, we map a rendered page image into a vocabulary-indexed sparse vector by applying an LM-head projection and SPLADE activation to the visual hidden states produced by the vision-language backbone. Given visual hidden states $\{\mathbf{h}_t\}_{t=1}^{L}$, the image-side sparse representation is
\begin{equation}
\mathbf{w}_p
=
\max_{t=1}^{L}
\log \left(1 + \mathrm{ReLU}(\mathrm{LMHead}(\mathbf{h}_{t}))\right)
\in \mathbb{R}^{|V|}.
\end{equation}
Here, $|V|$ is the vocabulary size and $\mathbf{w}_p[v]$ is the sparse weight assigned to vocabulary token $v$. The transformation applies ReLU activation followed by max pooling over visual tokens.

\paragraph{Query-encoding-free representation.}
\method{} avoids neural query encoding by adapting Li-LSR-style~\cite{nardini2025lilsr} learned query-token lookup to MMLSR. The query is represented as a weighted BoW vector, where each query token receives a learned vocabulary-level weight from a lookup table.
For each vocabulary token $v$, we compute
\begin{equation}
    a_v = \mathrm{softplus}(\mathbf{e}_v^\top \mathbf{u} + b),
\end{equation}
where $\mathbf{e}_v \in \mathbb{R}^{d}$ is the frozen token embedding, and $\mathbf{u} \in \mathbb{R}^{d}$ and $b$ are learned parameters. Given the BoW mask $m_q[v] \in \{0,1\}$ of query $q$, the query representation is
\begin{equation}
    \mathbf{w}_q[v] = m_q[v] \cdot a_v.
\end{equation}
After training, all $a_v$ values are stored in a lookup table, so inference only requires tokenization and weight lookup.

We make one important adaptation for the visual sparse setting. The original ReLU-based lookup in Li-LSR is prone to all-zero activation collapse when combined with visual sparse training and sparsity regularization, as shown in Section~\ref{sec:design_choices}. We therefore replace ReLU with softplus. This adaptation preserves the simplicity of Li-LSR-style lookup while improving training stability in the visual sparse setting.

\paragraph{Retrieval scoring.}
Retrieval is performed by sparse lexical matching. Given a query representation $\mathbf{w}_q \in \mathbb{R}^{|V|}$ and a document representation $\mathbf{w}_p \in \mathbb{R}^{|V|}$, the score is
\begin{equation}
\label{eq:scoring}
s(q,p)
=
\mathbf{w}_q^\top \mathbf{w}_p
=
\sum_{v \in V} \mathbf{w}_q[v]\mathbf{w}_p[v].
\end{equation}
The ranking loss based on Eq.~\ref{eq:scoring} and the sparsity regularizer for controlling active vocabulary dimensions are defined in Section~\ref{sec:training_objective}.

\subsection{Lexical Grounding through Caption-Gated Token Supervision}
\label{sec:caption_supervision}

The lexical grounding problem makes visual sparse training difficult: the top image-side activations provide only partial coverage of the document's actual lexical content. Query--document matching can then rely on generic or spurious dimensions, producing coarse ranking feedback. Caption-gated token supervision addresses this by using captions to selectively reinforce content-bearing lexical dimensions on the image side. As supported by the token-level analysis in Section~\ref{sec:design_choices}, this allows ranking loss to further activate retrieval-relevant dimensions that contain lexical information in the document, while generic or unsupported activations are left to be suppressed by the sparsity regularizer.

\paragraph{Training-time captions.}
We generate one offline caption for each training document and use captions only during training. Although caption quality may vary with alternative captioning prompts or pipelines, our focus is not caption generation itself, but how to use available captions as sparse supervision. We therefore follow the standard ColPali~\cite{faysse2025colpali} setup: captions are generated by a multimodal LLM with the ColPali prompt without modification; full prompt and generation details are provided in Section~\ref{sec:setup_training_data}. Because raw captions are not directly optimized for retrieval, we next convert them into retrieval-aware sparse representations.

\paragraph{Retrieval-aware caption representation.}
To obtain a retrieval-aware caption representation, we encode the caption with the same SPLADE-style sparse encoder used for the image side and apply the caption BoW mask to produce $\mathbf{w}_c \in \mathbb{R}^{|V|}$.

We train the caption representation with a query--caption ranking loss:
\begin{equation}
\label{eq:cap_rank}
    \mathcal{L}_{\mathrm{cap\_rank}}
    =
    -\frac{1}{B}
    \sum_{i=1}^{B}
    \log
    \frac{
        \exp(s(q_i,c_i)/\tau)
    }{
        \sum_{j=1}^{B} \exp(s(q_i,c_j)/\tau)
    },
\end{equation}
where $B$ is the batch size, $s(q_i,c_j)=\mathbf{w}_{q_i}^{\top}\mathbf{w}_{c_j}$, and $\tau$ is the ranking temperature.
This ranking loss turns the caption sparse representation into a retrieval-aware lexical signal used for gating.

\paragraph{Caption-gated token supervision.}
Let $\mathbf{w}^{\mathrm{sg}}_p, \mathbf{w}^{\mathrm{sg}}_c \in \mathbb{R}^{|V|}$ be stop-gradient sparse representations of the page image and caption.
For each vocabulary token $v$, we define the overlap gate:
\begin{equation}
    o[v] = \mathbf{w}^{\mathrm{sg}}_p[v] \cdot \mathbf{w}^{\mathrm{sg}}_c[v].
\end{equation}

The gate therefore selects vocabulary dimensions that both the image and caption representations consider useful. We then obtain $\alpha[v]$ from $o[v]$ by temperature sharpening and L1 normalization over vocabulary dimensions:
\begin{equation}
\alpha[v]
=
\frac{o[v]^{\,1/\tau_{\mathrm{cap}}}}{\sum_{v'} o[v']^{\,1/\tau_{\mathrm{cap}}}},
\end{equation}
where $\tau_{\mathrm{cap}}$ controls the sharpness of the caption-gated supervision.

Let $z_p[v]$ be the image-side vocabulary logit before the SPLADE ReLU and log-saturation.
We apply caption-gated token supervision to this pre-activation logit so that gated dimensions receive direct gradients.
The caption-gated token supervision loss is defined as
\begin{equation}
\label{eq:cap_gated}
    \mathcal{L}_{\mathrm{cap\_gated}}
    =
    -\frac{1}{B}
    \sum_{i=1}^{B}
    \sum_v
    \alpha_i[v]\,
    \log \sigma(z_{p_i}[v]).
\end{equation}

% Table 1 placed here so it appears at top of page 4 (Experiments page)
\begin{table*}[t!]
\centering
\small
\setlength{\tabcolsep}{3pt}
\begin{tabular*}{\textwidth}{l @{\extracolsep{\fill}} cllcccccccc}
\toprule
\textbf{Model} & \textbf{Size} & \textbf{Late-Int.} & \textbf{Q-Enc} & \textbf{ViD-v1} & \textbf{ViD-v2} & \textbf{ViD-v3} & \textbf{VRG} & \textbf{VOD} & \textbf{IRP} & \textbf{Avg} & \textbf{FL} \\
\midrule
\multicolumn{12}{l}{\textit{$\leq$ 1B Parameters: target serving regime comparison}} \\
ColFlor \cite{masry2025colflor}         & 0.17B & \checkmark & \checkmark & 77.4 & 43.1 & 36.8 & 68.0 & 57.3 & 52.5 & 55.8 & - \\
ColModernVBERT \cite{teiletche2025modernvbert} & 0.25B & \checkmark & \checkmark & \textbf{83.9} & \textbf{56.0} & \textbf{44.6} & \textbf{79.6} & \textbf{66.1} & \textbf{62.6} & \textbf{65.5} & - \\
Jina CLIP v2 \cite{koukounas2024jina}   & 0.9B &  & \checkmark & 55.7 & 28.5 & 25.7 & 48.1 & 47.2 & 26.6 & 38.6 & - \\
SigLIP2-L \cite{tschannen2025siglip2}   & 0.9B &  & \checkmark & 42.7 & 27.0 & 22.9 & 42.7 & 34.8 & 13.0 & 30.5 & - \\
BiModernVBERT \cite{teiletche2025modernvbert} & 0.25B &  & \checkmark & 67.6 & 35.7 & 28.9 & 60.5 & 53.4 & 31.8 & 46.3 & - \\
D2S \cite{nguyen2024expansion}          & 0.25B &  & \checkmark & 5.8 & 4.6 & 1.8 & 8.3 & 9.6 & 0.6 & 5.1 & 14.7 \\
JSDO-Sparse \cite{song2025dense_sparse} & 0.25B &  & \checkmark & 16.8 & 12.4 & 5.8 & 19.6 & 29.7 & 2.6 & 14.5 & 1814 \\
\method{} \textit{quality}             & 0.25B &  & & \textbf{77.4} & \textbf{49.9} & \textbf{40.9} & \textbf{76.4} & \textbf{61.7} & \textbf{54.0} & \textbf{60.1} & \textbf{1.9} \\
\method{} \textit{efficient}           & 0.25B &  & & 74.6 & 46.6 & 37.6 & 73.0 & 59.5 & 47.1 & 56.4 & 1.1 \\
\midrule
\multicolumn{12}{l}{\textit{Text-based Retrieval}} \\
BM25 {\small(caption \cite{qwenteam2025qwen3vl})}            & --- &  & & 67.5 & 44.1 & 38.3 & 76.5 & 58.0 & 38.4 & 53.8 & 1.3 \\
BM25 {\small(unstr.\ \cite{unstructured2024})}             & --- &  & & 68.2 & 41.7 & 38.7 & 61.1 & 51.2 & 65.7 & 54.4 & 1.4 \\
BGE-M3 {\small(caption)} \cite{chen2024bge}    & 0.57B &  & \checkmark & 64.8 & 53.5 & 38.8 & 72.0 & 60.7 & 37.0 & 54.5 & - \\
BGE-M3 {\small(unstr.)} \cite{chen2024bge}     & 0.57B &  & \checkmark & 63.0 & 50.9 & 45.0 & 55.8 & 52.1 & 55.7 & 53.8 & - \\
\midrule
\multicolumn{12}{l}{\textit{Fusion}} \\
\method{} + BiModernVBERT                    & 0.25/0.25B &  & \checkmark & 80.3 & 51.9 & 43.6 & 78.1 & 63.8 & 55.7 & 62.2 & - \\
\method{} + BM25 {\small(unstr.)}              & 0.25B &  & & 78.7 & 51.5 & 44.8 & 75.9 & 62.2 & 64.6 & 63.0 & - \\
\method{} + BiModernVBERT + BM25 {\small(unstr.)} & 0.25/0.25B &  & \checkmark & 80.7 & 53.1 & 47.2 & 77.9 & 64.0 & 64.1 & 64.5 & - \\
\midrule[\heavyrulewidth]
\multicolumn{12}{l}{\textit{$\geq$ 1B Parameters: high-capacity reference models}} \\
ColPali \cite{faysse2025colpali}        & 3B & \checkmark & \checkmark & 83.1 & 54.3 & 47.6 & 80.1 & 66.5 & 58.6 & 65.0 & - \\
NemColEmbed V2 8B \cite{moreira2026nemotron} & 8B & \checkmark & \checkmark & \textbf{91.9} & 65.9 & \textbf{64.2} & \textbf{90.7} & \textbf{73.5} & \textbf{74.8} & \textbf{76.8} & - \\
E5-V \cite{jiang2024e5v}               & 8B &  & \checkmark & 63.3 & 49.6 & 34.8 & 61.6 & 60.8 & 37.2 & 51.2 & - \\
VLM2Vec \cite{jiang2024vlm2vec}        & 4B &  & \checkmark & 49.2 & 41.5 & 26.3 & 52.1 & 54.0 & 17.1 & 40.0 & - \\
GME-Qwen2-7B \cite{zhang2025gme}       & 7B &  & \checkmark & \textbf{87.2} & \textbf{63.3} & \textbf{55.9} & \textbf{86.0} & \textbf{69.3} & \textbf{68.7} & \textbf{71.7} & - \\
Qwen3-VL-Emb-2B \cite{li2026qwen3vlemb} & 2B &  & \checkmark & 82.4 & 66.9 & 55.0 & 85.4 & 69.2 & 66.9 & 71.0 & - \\
\bottomrule
\end{tabular*}
\caption{NDCG@5 on six visual-document retrieval benchmarks: ViDoRe v1~\cite{faysse2025colpali}, v2~\cite{mace2025vidorev2}, and v3~\cite{loison2026vidorev3} (ViD-v1/v2/v3), VisRAG (VRG)~\cite{yu2025visrag}, VisDoc OOD (VOD)~\cite{meng2025vlm2vecv2}, and IRPAPERS (IRP)~\cite{shorten2026irpapers}. Avg is the mean over all six benchmarks; Late-Int. = late interaction; Q-Enc = neural query encoder required ($\checkmark$); FL = \flops{}. The best late-interaction and best single-vector model in each parameter-size band are bolded.}
\label{tab:main_results}
\end{table*}

\subsection{Training Objective and Deployment Path}
\label{sec:training_objective}

We train \method{} with three loss groups: ranking, caption-gated token supervision, and sparsity regularization.

For image-side ranking, we use an in-batch InfoNCE loss on the query--page score $s(q,p)$:
\begin{equation}
\mathcal{L}_{\mathrm{rank}}
=
-\frac{1}{B}
\sum_{i=1}^{B}
\log
\frac{
    \exp(s(q_i,p_i)/\tau)
}{
    \sum_{j=1}^{B}
    \exp(s(q_i,p_j)/\tau)
}.
\end{equation}
Here, $p_i$ is the positive page for $q_i$, other pages in the batch are negatives, and $\tau$ is the ranking temperature.
The caption-side ranking loss $\mathcal{L}_{\mathrm{cap\_rank}}$ has the same form with the query--caption score $s(q,c)$.

We control sparsity with the FLOPS regularizer~\cite{formal2021spladev2}.
For a batch of sparse representations, it is
\begin{equation}
\ell_{\mathrm{FLOPS}}
=
\sum_{j \in V}
\left(
\frac{1}{B}
\sum_{i=1}^{B}
\mathbf{w}_i[j]
\right)^2 .
\end{equation}
This penalizes vocabulary dimensions that are active across many examples.
We apply it separately to the image-side and caption-side batches, yielding $\ell_{\mathrm{FLOPS}}^{p}$ and $\ell_{\mathrm{FLOPS}}^{c}$, respectively.

We group the losses as
\begin{equation}
\begin{aligned}
\mathcal{L}_{\mathrm{R}}
&=
\mathcal{L}_{\mathrm{rank}}
+
\lambda_{\mathrm{cap\_rank}}
\mathcal{L}_{\mathrm{cap\_rank}},\\[6pt]
\mathcal{L}_{\mathrm{S}}
&=
\lambda_p
\ell_{\mathrm{FLOPS}}^{p}
+
\lambda_c
\ell_{\mathrm{FLOPS}}^{c},\\[6pt]
\mathcal{L}
&=
\mathcal{L}_{\mathrm{R}}
+
\lambda_{\mathrm{cap\_gated}}
\mathcal{L}_{\mathrm{cap\_gated}}
+
\mathcal{L}_{\mathrm{S}}.
\end{aligned}
\end{equation}
All $\lambda$ values are tunable hyperparameters.

At deployment, all caption-side components are removed.
Retrieval uses the sparse dot product between query and image-side document representations.

\section{Experiments}
\label{sec:experiments}

\subsection{Overview}
\label{sec:exp_overview}

We evaluate \method{} as a scalable lexical retriever for visual-documents.
We first report benchmark retrieval quality with indexing throughput, then test production-scale retrieval on an 18.7M-document corpus, measuring serving latency, scaling robustness, and dense-retriever complementarity.
Finally, we ablate caption-gated token supervision and analyze its token-level effects.

\subsection{Setup}

\paragraph{Backbone and training data.}
\label{sec:setup_training_data}
We use ModernVBERT \cite{teiletche2025modernvbert} ($\sim$250M parameters), a vision-language backbone with a SigLIP-2 vision encoder, ModernBERT text components, and an LM head.
We choose this backbone for its compact size (suitable for large-scale indexing) and its built-in LM head that naturally projects to vocabulary space.
We use the same training data as BiModernVBERT~\cite{teiletche2025modernvbert}, the visual-document retrieval adaptation of ModernVBERT. The training mixture consists of ColPali training data (118K image--query pairs) mixed with RLHN~\cite{thakur2025rlhn} (300K text retrieval pairs, 2 hard negatives per query) at a 3:1 text-to-image ratio.
BiModernVBERT additionally uses hard negatives on the image side, but the mined hard negatives are not publicly available, so for image--query pairs we rely on in-batch negatives only.
We adopt the BiModernVBERT~\cite{teiletche2025modernvbert} training recipe, which mixes image and text retrieval data.
Captions are generated offline only for the ColPali image--query pairs (text retrieval pairs are unchanged) using Qwen3-VL-30B (228 median words per caption) with the ColPali prompt without modification.
The full prompt is:
\begin{quote}
\small
You are an assistant specialized in document analysis. Given a table or a figure, provide a detailed summary (maximum 3000 characters). Your summary should be qualitative and not quantitative. Here is the table/figure: Answer ONLY with the caption.
\end{quote}

\paragraph{Hyperparameters.}
We train for 3 epochs on 4$\times$H100 GPUs with batch size 42 per GPU using AdamW~\cite{loshchilov2019adamw} with a WSD schedule (5\% warmup, 20\% decay) and learning rate $5{\times}10^{-4}$.
LoRA~\cite{hu2022lora} ($r{=}32$) is applied to both the encoder and the LM head; document tokens are aggregated via max-pooling.
The ranking softmax temperature $\tau$ and caption-gated focus temperature $\tau_{\mathrm{cap}}$ are set to 0.1 and 0.5, respectively. Sparsity-regularizer warmup is 500 steps, and the caption-gated token supervision weight is $\lambda_{\mathrm{cap\_gated}}{=}5$ with caption-side FLOPs weight $\lambda_c{=}0.005$.
The \emph{quality} and \emph{efficient} operating points share all settings above and differ only in two regularizers: passage FLOPs weight ($\lambda_p{=}0.01$ for quality vs.\ $0.05$ for efficient) and caption sparse-rank weight ($\lambda_{\mathrm{cap\_rank}}{=}1.0$ vs.\ $0.5$).

\paragraph{Evaluation.}
We evaluate on standard visual-document retrieval benchmarks, focusing on English-language settings: \textbf{ViDoRe} \textbf{v1}~\cite{faysse2025colpali}, \textbf{v2}~\cite{mace2025vidorev2}, and \textbf{v3}~\cite{loison2026vidorev3} (\textbf{ViD-v1/v2/v3}), \textbf{VisRAG} (\textbf{VRG})~\cite{yu2025visrag}, \textbf{VisDoc OOD} (\textbf{VOD})~\cite{meng2025vlm2vecv2}, and \textbf{IRPAPERS} (\textbf{IRP})~\cite{shorten2026irpapers}.
Together, these six benchmarks cover a comprehensive range of domains---from arXiv papers~\cite{faysse2025colpali,shorten2026irpapers} to financial and enterprise reports~\cite{faysse2025colpali,meng2025vlm2vecv2}---and retrieval situations---from short keyword queries (\textit{``two factor authentication vs single factor security mechanisms''})~\cite{faysse2025colpali} to figure-grounded queries with LaTeX symbols (\textit{``value of $E_{\alpha}(k)$''})~\cite{yu2025visrag}.
We report NDCG@5~\cite{jarvelin2002ndcg} unless stated otherwise.
For lexical sparse retrievers, including MMLSR and BM25-based systems, we report \flops{}~\cite{formal2021splade} as an efficiency proxy for query--document sparse matching. It estimates the average number of floating-point operations from overlapping active vocabulary dimensions in query--document sparse matching. We abbreviate this metric as FL in tables.
Retrieval wall time is measured on a 2-socket Intel Xeon Platinum 8462Y+ server; each latency table reports the CPU thread count used for the corresponding measurement.
Document and query encoding are measured on a single NVIDIA H100 GPU.

\subsection{Benchmark Retrieval Quality}
\label{sec:main_results}

Table~\ref{tab:main_results} compares \method{} with baselines in our target serving regime, along with high-capacity reference models.

\paragraph{Main result.}
\method{} outperforms the two most relevant baselines in this regime.
BiModernVBERT serves as the same-scale state-of-the-art compact dense baseline for visual-document retrieval, with fast query encoding and retrieval.
BM25 over OCR text or VLM-generated captions represents the query-encoding-free lexical regime.
The OCR baseline uses \texttt{unstructured}~\cite{unstructured2024}, a widely used document parsing pipeline, with Tesseract~\cite{smith2007tesseract} in the \texttt{hi\_res} setting, following ColPali.
The caption baseline uses Qwen3-VL-30B-A3B with the ColPali prompt.
For both text sources, we build BM25 indexes with Pyserini~\cite{lin2021pyserini}.
\method{} offers two operating points for inverted-index retrieval: the \emph{quality} variant maximizes retrieval accuracy at \flops{} = 1.9, while the \emph{efficient} variant reduces retrieval cost to \flops{} = 1.1.
On the six-benchmark average, they reach 60.1 and 56.4 NDCG@5, respectively, and both outperform the main target-regime baselines.
Compared with BiModernVBERT's 46.3---trained on the same backbone and the same training data---they improve by +13.8pp and +10.1pp; they also improve over OCR-based BM25 (54.4) and caption-based BM25 (53.8). We include prior MMLSR methods, D2S~\cite{nguyen2024expansion} and JSDO-Sparse~\cite{song2025dense_sparse}, as reference points rather than directly comparable baselines, since they were not trained on visual-document retrieval data. Their results illustrate that existing MMLSR methods do not yet cover the visual-document retrieval task.

\paragraph{Complementarity with dense retrieval.}
\method{} is not only a standalone lexical retriever; it can also be added to an existing dense retriever or BM25 system as a complementary signal.
We use Relative Score Fusion (RSF)~\cite{bruch2023fusion}: scores from each retriever are min--max normalized per query and then combined with fusion weights that sum to 1, where $w_{\mathrm{V}}$, $w_{\mathrm{B}}$, and $w_{\mathrm{M}}$ denote the weights for \method{}, BiModernVBERT, and BM25, respectively.

With RSF, \method{} \emph{quality} plus BiModernVBERT, with $w_{\mathrm{V}}=0.6$ and $w_{\mathrm{B}}=0.4$, raises the average NDCG@5 from 60.1 to 62.2, indicating that the sparse and dense representations capture complementary evidence.
\method{} also complements text-derived lexical retrieval: fusing it with BM25 over unstructured text, with $w_{\mathrm{V}}=0.7$ and $w_{\mathrm{M}}=0.3$, improves the average NDCG@5 to 63.0.
This two-way lexical fusion requires no neural query encoding, making it useful when serving-time compute is highly constrained.
A three-way fusion of \method{}, BiModernVBERT, and BM25 reaches 64.5, with $w_{\mathrm{V}}=0.5$, $w_{\mathrm{B}}=0.3$, and $w_{\mathrm{M}}=0.2$.
These results suggest that \method{} can be added as a complementary lexical layer on top of either existing dense visual retrievers or OCR-based BM25 retrieval systems.

The fused same-scale systems also narrow the gap to stronger neural retrievers.
In particular, the three-way fusion comes within roughly one NDCG@5 point of ColModernVBERT, the same-scale late-interaction model, while avoiding multi-vector scoring in the first-stage retrievers.

\begin{figure}[H]
\centering
\includegraphics[width=\columnwidth]{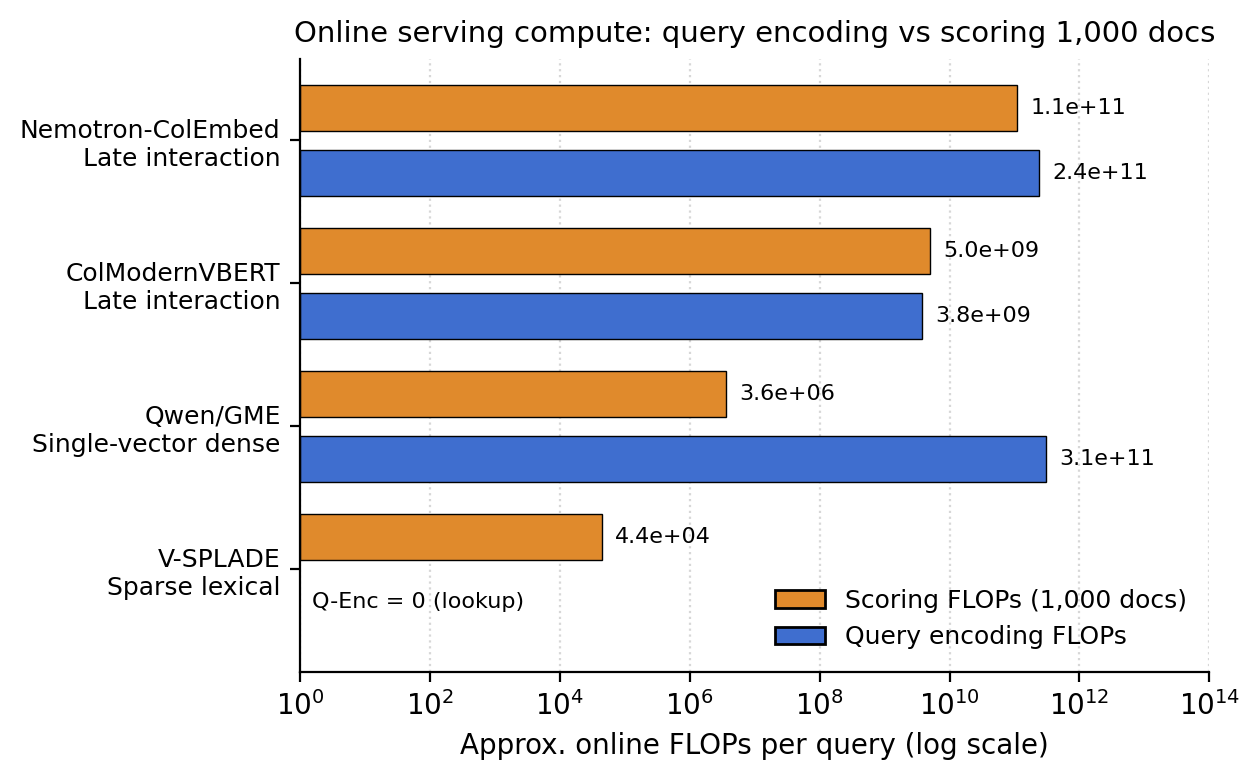}
\caption{Online serving FLOPs per query.}
\label{fig:flops_serving}
\end{figure}

\paragraph{Efficiency: }\method{} has substantially lower serving cost than high-capacity models from other retrieval regimes.
Although these models serve as references for the current accuracy upper bound, they do not operate in the same serving regime as \method{}.
To quantify this gap, we approximate query-time serving cost as the FLOPs required to score 1{,}000 documents for a single query, averaged over 50 sampled queries and 1{,}000 sampled documents from the six benchmarks.
We measure query-encoding FLOPs with \texttt{fvcore}~\cite{fvcore} under each model's actual evaluation template.
We compute scoring FLOPs analytically from each retrieval form: sparse dot products for \method{}, dense vector products for single-vector retrievers, and MaxSim operations for multi-vector retrievers.
Figure~\ref{fig:flops_serving} shows that the strongest models in the competing dense and late-interaction regimes require higher online compute than \method{}.

\paragraph{Document encoding throughput.}
We measure document encoding throughput to test whether \method{} scales efficiently compared with other query-encoding-free lexical candidates.
We compare \method{} against two text-generation/extraction pipelines used to build BM25-style lexical indexes: a Qwen3-VL-30B-A3B MoE caption generator (3B effective active parameters at inference, served with vLLM~\cite{kwon2023vllm}) and an OCR pipeline (\texttt{unstructured}~\cite{unstructured2024} with Tesseract~\cite{smith2007tesseract} \texttt{hi\_res}).
For \method{}, we run inference with a standard PyTorch \texttt{DataLoader}; the caption and OCR baselines are run with their respective serving stacks.
All systems are measured on a single H100 GPU with 4 CPU cores, using 1{,}000 sampled documents from the six benchmarks.
As shown in Table~\ref{tab:encoding_speed}, \method{} encodes page images at 20.19 pages/sec, compared with 0.83 pages/sec for caption generation and 0.90 pages/sec for the OCR pipeline, making it over $20\times$ faster than both text-extraction alternatives.

\begin{table}[H]
\centering
\small
\setlength{\tabcolsep}{4pt}
\begin{tabular}{lccc}
\toprule
\textbf{Method} & \textbf{\method{}} & \textbf{Caption gen.} & \textbf{OCR pipeline} \\
\midrule
Pages/sec & \textbf{20.19} & 0.83 & 0.90 \\
\bottomrule
\end{tabular}
\caption{Document encoding throughput.}
\label{tab:encoding_speed}
\end{table}

Taken together, these results show that \method{} is competitive standalone, complementary to OCR-based BM25 and BiModernVBERT, and efficient in both online serving and document encoding.

\subsection{Production-Scale Retrieval}
\label{sec:large_scale}

The standard benchmarks in Section~\ref{sec:main_results} contain under 25{,}000 pages per benchmark, which is still far from production-scale retrieval.
We therefore evaluate \method{} on an 18.7M-page corpus built from PDFA~\cite{montalvo2024pdfa} and DocMatix~\cite{laurencon2024idefics3}.
PDFA is a large open-source PDF corpus derived from SafeDocs~\cite{safedocs2021}, while DocMatix is a labeled subset of PDFA with generated question--answer pairs.
We render PDFA into page images and treat each page image as a retrievable item.
Each DocMatix question is then converted into a recall query whose relevant items are the page images from the corresponding source document.

We compare \method{} with BiModernVBERT, the same-scale compact state-of-the-art dense retriever trained for visual-document retrieval. For compactness, we denote BiModernVBERT as \textit{vbert} in the tables.
Dense retrieval uses FAISS~\cite{johnson2017faiss} Flat and HNSW~\cite{malkov2020hnsw}, while \method{} uses PISA~\cite{pisa2019} inverted-index retrieval.
For HNSW indexing, we use $M{=}32$ and $\texttt{efConstruction}{=}128$.
We report the best latency--recall trade-off at $\texttt{efSearch}{=}256$.
All recall and latency measurements in Table~\ref{tab:five_way} are computed on a fixed set of 1{,}000 randomly sampled queries.
For \method{}, we evaluate both full-index search and a two-stage sparse search.
Full search scores the complete inverted index.
The two-stage variant adopts the idea of Two-Step SPLADE~\cite{lassance2024twostepsplade}: a pruned index is used for fast candidate generation, and the selected candidates are rescored with the full index.
In our implementation, the first stage uses a pruned top-50 index.
Unless stated otherwise, we use the \emph{quality} variant of \method{} in the production-scale retrieval experiment; the \emph{efficient} variant is used to show the latency--accuracy trade-off.

\begin{table}[h!]
\centering
\small
\setlength{\tabcolsep}{4pt}
\begin{tabular}{lccc}
\toprule
\textbf{Method} & \textbf{R@5 / R@100} & \textbf{q\_enc (cpu/gpu)} & \textbf{ms/q (j=1 / 20)} \\
\midrule
vbert Flat  & 0.090 / 0.228 & 85.87 / 0.08 & 249.55 / 47.31 \\
vbert HNSW  & 0.071 / 0.191 & 85.87 / 0.08 & 3.63 / 0.31 \\
\midrule
quality Full         & 0.228 / 0.396 & --- & 59.25 / 5.89 \\
quality Two-Stage    & 0.183 / 0.278 & --- & 4.22 / 0.45 \\
efficient Full       & 0.202 / 0.365 & --- & 25.14 / 2.36 \\
efficient Two-Stage  & 0.183 / 0.290 & --- & 3.78 / 0.38 \\
\bottomrule
\end{tabular}
\caption{Recall and latency on the 18.7M-document corpus.}
\label{tab:five_way}
\end{table}

\paragraph{Retrieval quality and latency at 18.7M scale.}
\method{} consistently outperforms the same-backbone dense baseline on the 18.7M-page corpus without neural query encoding.
Table~\ref{tab:five_way} shows a clear gap at the top ranks: full-index \method{} reaches R@5=0.228, compared with 0.090 for FAISS Flat BiModernVBERT.
The same trend holds at a broader cutoff, where full-index \method{} reaches R@100=0.396, compared with 0.228 for Flat.
The two-stage sparse variant, which uses a pruned first-stage index before rescoring, reaches R@5=0.183 and R@100=0.278, well above the dense HNSW baseline at R@5=0.071 and R@100=0.191.

Table~\ref{tab:five_way} shows that \method{} reaches the low-latency regime without neural query encoding.
In the low-latency setting, two-stage sparse retrieval reaches 0.45 ms/query for the quality model and 0.38 ms/query for the efficient model with 20 CPU threads, which is comparable to HNSW retrieval at 0.31 ms/query.
Unlike the dense baselines, these sparse variants require no neural query encoding, avoiding the additional 85.87 ms CPU or 0.08 ms batched query-encoding cost on an H100 GPU.
When recall is prioritized, full-index sparse retrieval provides a higher-recall operating point.
Although slower than HNSW, the quality variant still runs in the sub-10 ms regime at 5.89 ms/query with 20 CPU threads and achieves the highest recall among the same-backbone systems.

\paragraph{Larger gains on lexically specific queries.}
\method{} shows its largest gains on queries when exact lexical evidence is likely to matter.
We split queries by two heuristics---whether they contain digits or uppercase characters beyond the first character---which approximate numerals and proper-noun-like expressions.
As shown in Table~\ref{tab:query_subset}, the sparse-over-dense gap is larger on these subsets than on all queries, peaking when both features are present: \method{} reaches R@5=0.363 versus 0.135 for the dense baseline.
This suggests that the production-scale gains are concentrated in lexically specific queries.

\begin{table}[H]
\centering
\setlength{\tabcolsep}{4pt}
\begin{tabular}{lrcccc}
\toprule
\textbf{Query Subset} & \textbf{\%} & \textbf{Dense} & \textbf{\method{}} & \textbf{Gap} & \textbf{Gap/ALL} \\
\midrule
Has digit       & 17.7\% & 0.127 & 0.343 & +0.216 & 1.57$\times$ \\
Has uppercase   & 64.8\% & 0.115 & 0.306 & +0.191 & 1.38$\times$ \\
\textbf{Upper + Digit} & \textbf{15.2\%} & \textbf{0.135} & \textbf{0.363} & \textbf{+0.228} & \textbf{1.65$\times$} \\
\midrule
All             & 100\%  & 0.090 & 0.228 & +0.138 & 1.00$\times$ \\
\bottomrule
\end{tabular}
\caption{Query subset analysis on the 18.7M corpus (R@5).}
\label{tab:query_subset}
\end{table}

\begin{figure}[H]
\centering
\includegraphics[width=0.85\columnwidth]{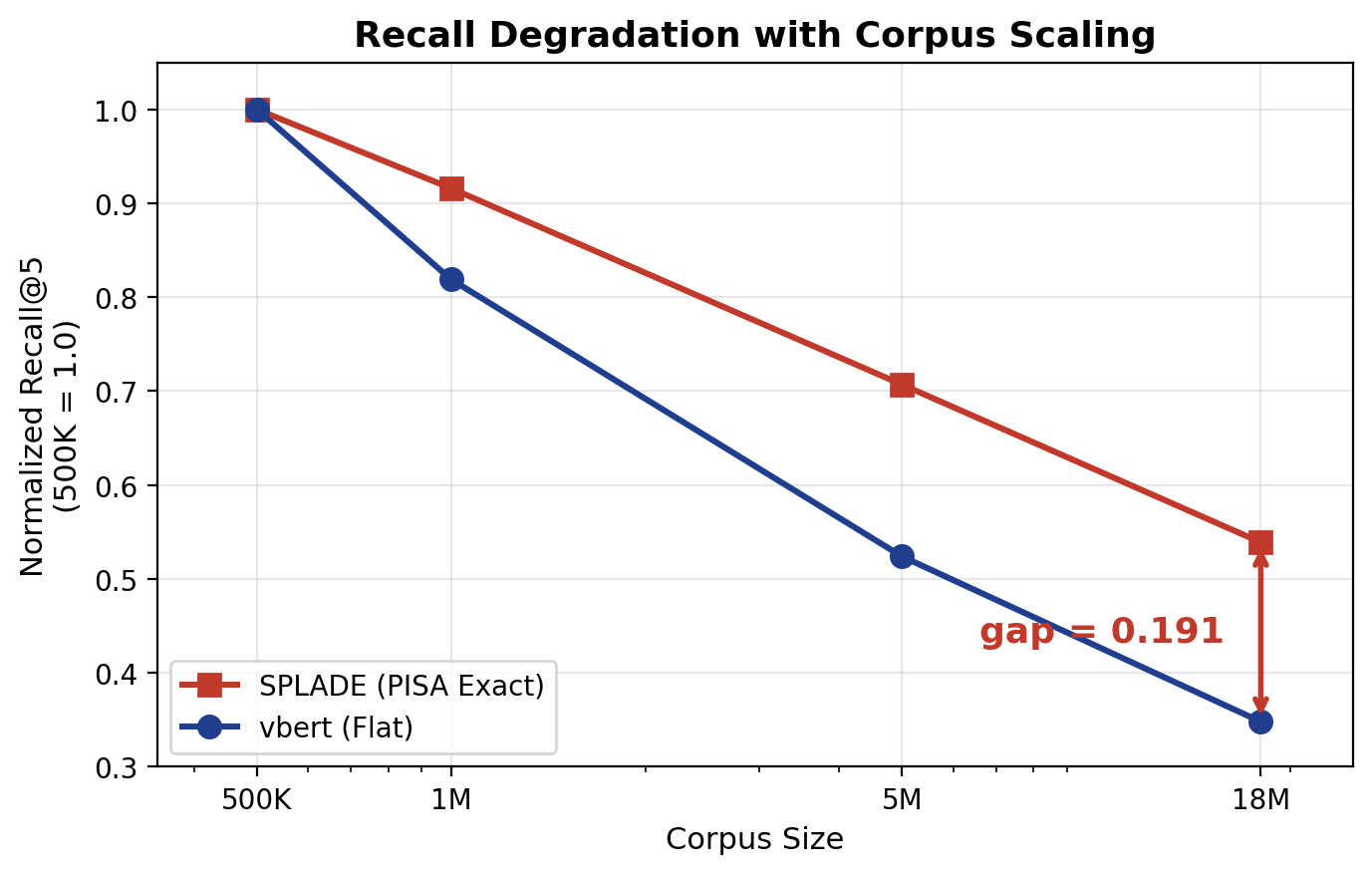}
\caption{Normalized recall degradation with corpus scaling (R@5, normalized to 500K~=~1.0).}
\label{fig:degradation}
\end{figure}

\paragraph{Robustness under corpus scaling.}
Sparse lexical representations may be more robust under corpus scaling because they operate over a much larger vocabulary-indexed space than fixed-dimensional dense embeddings. Motivated by recent work on the dimensional limits of dense retrieval capacity~\cite{weller2026theoretical}, we evaluate this hypothesis by scaling the corpus from 500K to 18.7M pages and report R@5/R@100 in Table~\ref{tab:scaling}.
At R@5, the dense retriever drops from 0.260 to 0.090, retaining only 35\% of its 500K performance, whereas \method{} drops from 0.424 to 0.228, retaining 54\%.
Figure~\ref{fig:degradation} visualizes the same trend with recall normalized to the 500K setting: the dense--sparse gap widens monotonically as the corpus grows.
This suggests that the lexical sparse representations are more robust to performance degradation under corpus scaling.

\begin{table}[H]
\centering
\small
\setlength{\tabcolsep}{5pt}
\begin{tabular}{lcccc}
\toprule
\textbf{Method} & \textbf{500K} & \textbf{1M} & \textbf{5M} & \textbf{18M} \\
\midrule
vbert             & 0.26/0.45 & 0.21/0.41 & 0.14/0.31 & 0.09/0.23 \\
\textbf{\method{}} & \textbf{0.42/0.60} & \textbf{0.39/0.56} & \textbf{0.30/0.47} & \textbf{0.23/0.40} \\
\bottomrule
\end{tabular}
\caption{Recall (R@5 / R@100) vs.\ corpus scale.}
\label{tab:scaling}
\end{table}

\paragraph{Complementarity with existing neural retrievers at large scale.}
\method{} remains complementary to existing neural retrievers at production scale, including a billion-scale dense model and a late-interaction retriever. Dense and late-interaction retrievers can also be practical when sufficient GPU resources are available, especially for compact models such as BiModernVBERT. We therefore ask whether \method{} adds value beyond replacing dense retrieval. On the 18.7M corpus, we evaluate two integration scenarios: score fusion with dense retrievers and first-stage retrieval for multi-vector reranking.

For score fusion, we use Relative Score Fusion (RSF)~\cite{bruch2023fusion}, where scores from each retriever are min--max normalized per query and combined as
$s_{\mathrm{fused}} = w_d \cdot s_{\mathrm{sparse}} + (1 - w_d) \cdot s_{\mathrm{dense}}$,
where $w_d$ denotes the mixing weight assigned to \method{}.
We test fusion with two representative dense retrievers at different scales: Qwen3-VL-Embedding-2B~\cite{li2026qwen3vlemb}, a billion-scale SOTA dense retriever, and BiModernVBERT~\cite{teiletche2025modernvbert}, a compact same-backbone dense retriever. We denote them as \textit{qwen} and \textit{vbert} in Table~\ref{tab:rsf_p44}.
Table~\ref{tab:rsf_p44} shows that \method{} improves both dense systems.
With Qwen3-VL-Embedding-2B, fusion improves R@5 from 0.327 to 0.343.
With BiModernVBERT, fusion improves R@5 from the better standalone system by +2.4pp.
By contrast, RSF between the two dense retrievers does not improve over the stronger dense retriever, suggesting that the gains are unlikely to be explained solely by ensembling retrievers.
These gains suggest that the lexical signal from \method{} can complement dense retrievers across model scales.

\begin{table}[h!]
\centering
\setlength{\tabcolsep}{4pt}
\begin{tabular}{llcccc}
\toprule
\textbf{Fusion} & \textbf{K} & \textbf{Sparse} & \textbf{Dense} & \textbf{Best Fused} & \textbf{$w_d^\star$} \\
\midrule
\multirow{2}{*}{+ qwen}
 & R@5   & 0.228 & 0.327 & \textbf{0.343} \small{(+1.6pp)} & \multirow{2}{*}{0.2} \\
 & R@100 & 0.396 & 0.491 & \textbf{0.503} \small{(+1.2pp)} &                       \\
\midrule
\multirow{2}{*}{+ vbert}
 & R@5   & 0.228 & 0.090 & \textbf{0.252} \small{(+2.4pp)} & \multirow{2}{*}{0.8} \\
 & R@100 & 0.396 & 0.228 & \textbf{0.405} \small{(+0.9pp)} &                       \\
\bottomrule
\end{tabular}
\caption{Relative Score Fusion (RSF) on the 18.7M-document corpus.}
\label{tab:rsf_p44}
\end{table}

\begin{figure}[h!]
\centering
\includegraphics[width=0.85\columnwidth]{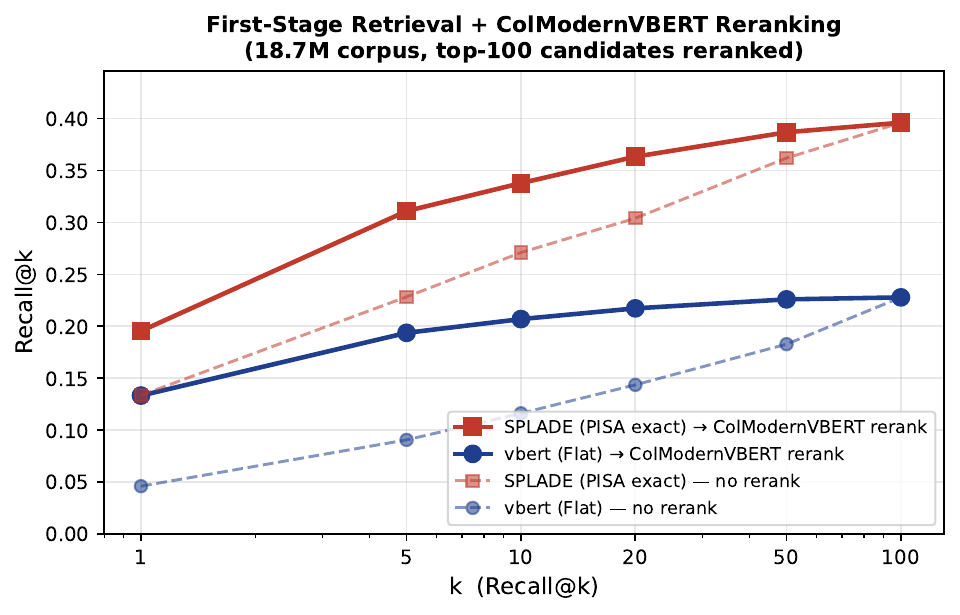}
\caption{Two-stage retrieval with ColModernVBERT reranking.}
\label{fig:rerank}
\end{figure}

We also test whether \method{} can improve a two-stage pipeline with a stronger multi-vector reranker.
Multi-vector visual retrievers are accurate but too expensive to scan the full 18.7M corpus, so they require a cheap first-stage retriever.
Figure~\ref{fig:rerank} compares two top-100 first-stage pipelines before ColModernVBERT reranking: BiModernVBERT $\rightarrow$ ColModernVBERT and \method{} $\rightarrow$ ColModernVBERT.

After reranking, the \method{} first-stage pipeline reaches 0.311 R@5, up from 0.228 before reranking, whereas the dense first-stage pipeline reaches 0.194, up from 0.090.
The difference is governed by the first-stage recall ceiling: \method{} retrieves a stronger top-100 candidate set than BiModernVBERT, with R@100 of 0.396 versus 0.228.
Thus, \method{} is not only an efficient standalone retriever, but also a complementary lexical component that boosts dense retrievers through score fusion and provides stronger candidates for multi-vector reranking.

\medskip

Overall, these results show that \method{} is effective at large corpus scale, especially for lexically specific queries. \method{} also scales robustly and complements stronger dense and late-interaction systems as a lexical layer.

\subsection{Ablations and Analysis}
\label{sec:c1_ablation}
\label{sec:design_choices}
\label{sec:coverage_analysis}

We analyze caption-gated token supervision from three perspectives: whether the signal is necessary, whether our design choices are effective, and how the signal acts on the image sparse representation during training. Through component ablations and alternative-design studies, we first examine the contribution of caption-gated token supervision to the quality--efficiency trade-off. We then analyze token-level training dynamics to show how this signal complements and reshapes the visual passage sparse embedding.
We use the efficient \method{} variant ($\lambda_p{=}0.05$) and run quantitative ablations across all six benchmarks.

\begin{table}[h!]
\centering
\setlength{\tabcolsep}{4pt}
\resizebox{\columnwidth}{!}{%
\begin{tabular}{lcc}
\toprule
\textbf{Variant} & \textbf{NDCG@5} & \textbf{\flops{}} \\
\midrule
Baseline (BoW)            & .537 & 1.88 \\
+Li-LSR                   & .549 & 3.06 \\
\textbf{+Li-LSR + cap-gated loss (Eq.~\ref{eq:cap_gated})} & \textbf{.564} & \textbf{1.10} \\
\midrule
\multicolumn{3}{l}{\textit{Alternative design choices}} \\
+Li-LSR + cap rank loss (Eq.~\ref{eq:cap_rank})              & .556 & 2.53 \\
+Li-LSR + cap cos-sim loss + cap rank loss (Eq.~\ref{eq:cap_rank}) & .014\textsuperscript{$\dagger$} & 0.00 \\
+Li-LSR (ReLU) + cap-gated loss (Eq.~\ref{eq:cap_gated})     & .014\textsuperscript{$\dagger$} & 0.00 \\
\bottomrule
\end{tabular}%
}
\caption{Component ablation on \method{} \textit{efficient}. $\dagger$~marks runs where training collapses.}
\label{tab:component_ablation}
\end{table}

\paragraph{Ablating the training signal.}
Table~\ref{tab:component_ablation} shows that caption-gated token supervision is the key component behind the improved quality--efficiency trade-off.
Li-LSR alone improves over the binary BoW baseline (.537$\rightarrow$.549), but increases \flops{} from 1.88 to 3.06.
With caption-gated supervision, NDCG@5 rises to .564 and \flops{} drops to 1.10.
Compared with caption-based BM25 baselines, Li-LSR alone reaches a similar quality range; caption-gated supervision opens a clear gap, improving NDCG@5 by more than two points while reducing \flops{}.

The alternative-design variants in Table~\ref{tab:component_ablation} further support our design choices.
Treating captions only as additional ranking positives provides a modest improvement over Li-LSR alone (.549$\rightarrow$.556), while also slightly reducing FLOPs (3.06$\rightarrow$2.53).
Directly aligning passage and caption sparse vectors with cosine similarity was unstable: it collapsed under the efficient-variant hyperparameters across multiple seeds, and even with the quality-variant hyperparameters it reached only 56.16 NDCG@5 at 2.47 FLOPs, below the original quality model's 60.1 NDCG@5 at 1.9 FLOPs.
Replacing the softplus Li-LSR activation with ReLU consistently collapsed training across multiple seeds, including runs with the quality-variant hyperparameters.

\paragraph{How lexical grounding reshapes sparse activations during training.}
Finally, we examine what additional lexical evidence the caption sparse embedding provides beyond the visual passage embedding alone.
For this analysis, we train \method{} for one epoch on 90\% of the ColPali image--query pairs, and then inspect the held-out 10\%.
On these unseen samples, we compare the top sparse tokens produced by the visual passage encoder and by the corresponding caption encoder.

\begin{table}[h!]
\centering
\small
\setlength{\tabcolsep}{4pt}
\begin{tabular}{lcc}
\toprule
\textbf{Retrieval source} & \textbf{Hit@1} & \textbf{Hits / 1000} \\
\midrule
Passage-only (image) & 0.43 & 427 \\
Caption-only (text)  & 0.44 & 440 \\
\textbf{Union (max)} & \textbf{0.53} & \textbf{530} \\
\bottomrule
\end{tabular}
\caption{Union analysis on 1000 unseen samples (top-30 sparse tokens).}
\label{tab:token_synergy_hit}
\end{table}

The passage and caption sparse embeddings do not activate the same vocabulary dimensions: 64\% of top-30 tokens are disjoint on average.
Table~\ref{tab:token_synergy_hit} shows that this complementarity is useful for retrieval: passage-only and caption-only tokens reach Hit@1 of 0.43 and 0.44, while taking their union improves Hit@1 to 0.53.
This indicates that passage and caption embeddings provide different views of the same visual-document rather than redundant token sets.

\begin{table}[H]
\centering
\small
\setlength{\tabcolsep}{2pt}
\begin{tabular}{@{}lp{0.95\columnwidth}@{}}
\toprule
\multicolumn{2}{l}{\textbf{Case 1}} \\
Q & \textbf{SVM}~{\scriptsize\textbf{(0.77)}}, score~{\scriptsize(0.67)}, figure~{\scriptsize(0.66)}, represent~{\scriptsize(0.64)}, given~{\scriptsize(0.60)} \\
P & Race~{\scriptsize(1.55)}, race~{\scriptsize(1.49)}, White~{\scriptsize(1.37)}, score~{\scriptsize(1.25)}, Racing~{\scriptsize(1.23)} \\
P* & \textbf{SVM}~{\scriptsize\textbf{(0.18)}} \\
C & \textbf{SVM}~{\scriptsize\textbf{(1.02)}}, VM~{\scriptsize(0.87)}, trend~{\scriptsize(0.73)}, intersection~{\scriptsize(0.71)}, divergence~{\scriptsize(0.70)} \\
\midrule
\multicolumn{2}{l}{\textbf{Case 2}} \\
Q & \textbf{ECG}~{\scriptsize\textbf{(0.80)}}, illustrated~{\scriptsize(0.70)}, changes~{\scriptsize(0.69)}, image~{\scriptsize(0.66)}, specific~{\scriptsize(0.61)} \\
P & Wave~{\scriptsize(1.49)}, wave~{\scriptsize(1.39)}, Wave~{\scriptsize(1.36)}, cardia~{\scriptsize(1.25)}, age~{\scriptsize(1.19)} \\
P* & \textbf{ECG}~{\scriptsize\textbf{(0.13)}} \\
C & \textbf{ECG}~{\scriptsize\textbf{(1.11)}}, tachy~{\scriptsize(0.97)}, cardi~{\scriptsize(0.93)}, tracing~{\scriptsize(0.88)}, heart~{\scriptsize(0.75)} \\
\bottomrule
\end{tabular}
\caption{Token synergy case studies.}
\label{tab:token_synergy}
\end{table}

Table~\ref{tab:token_synergy} qualitatively illustrates this effect.
Q, P, and C denote the top activated tokens from the query, image passage, and caption representations, while P* reports the image-passage activation of the top caption token.
Captions often highlight retrieval-relevant tokens that are weak in the passage embedding alone: in Case 1, ``SVM'' is strongly activated by the query and caption (0.77 and 1.02), but only weakly by the image passage (0.18). Caption-gated token supervision can therefore reinforce such under-activated image-side dimensions that the passage encoder would otherwise miss.

\begin{figure}[H]
\centering
\includegraphics[width=0.85\columnwidth]{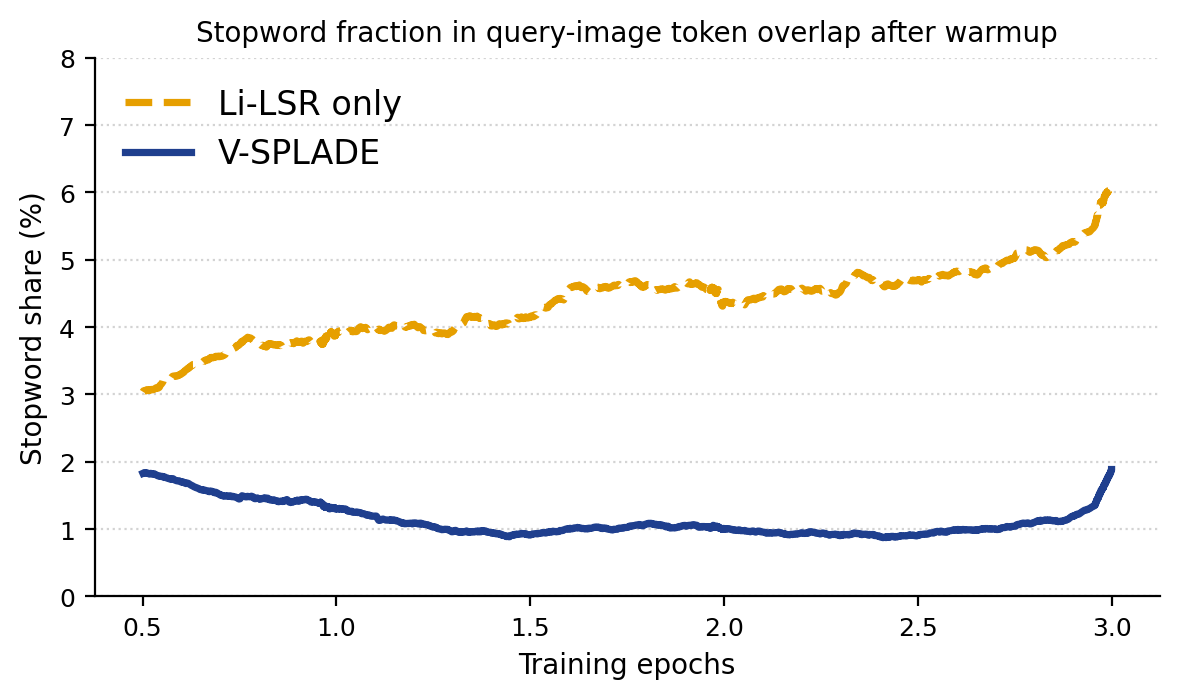}
\caption{Stopword fraction among query--image overlap.}
\label{fig:stopword_training}
\end{figure}

We next examine how this mechanism changes training dynamics. Figure~\ref{fig:stopword_training} tracks the stopword fraction among query--image matched tokens during training, using the 33 Lucene stopwords with casing variants~\cite{lin2021pyserini}. After the 0.5-epoch warmup, Li-LSR increasingly matches queries through stopwords, whereas \method{} keeps this fraction much lower. This suggests that caption-gated supervision shifts matching away from generic lexical dimensions toward content-bearing document tokens.

\begin{table}[h!]
\centering
\small
\setlength{\tabcolsep}{4pt}
\begin{tabular}{lcc}
\toprule
 & \textbf{Li-LSR only} & \textbf{V-SPLADE} \\
\midrule
50\% masked (NDCG@5) & $-$79.4\% & \textbf{$-$56.5\%} \\
Top-1 vs top-2 margin (mean) & 0.319 & \textbf{0.754} \small{($2.36\times$)} \\
Stopword tokens / passage & 6.95 & \textbf{1.69} \\
Active tokens / passage & 603 & \textbf{300} \\
\bottomrule
\end{tabular}
\caption{Masking robustness, ranking margin, stopword activations, and passage token list length.}
\label{tab:masking_margin_stopword}
\end{table}

Table~\ref{tab:masking_margin_stopword} summarizes the resulting token-level behavior. Compared with Li-LSR alone, \method{} is less sensitive to masking high-weight query tokens, yields a larger top-1/top-2 margin, and produces fewer stopword activations with shorter passage token lists. Together, these results suggest that caption-gated token supervision makes sparse representations more selective and less dominated by noisy or brittle lexical matches.

\paragraph{Effect of caption-generator scale.}
We also test whether the effectiveness of caption-gated token supervision depends on the caption generator scale. We generate captions with Qwen3-VL models of varying sizes~\cite{qwenteam2025qwen3vl} and train \method{} with each caption source under the \emph{efficient} variant hyperparameters. Across different caption generators, caption-gated supervision consistently improves retrieval quality over the Li-LSR only baseline, indicating that the gain is not tied to a single caption model but comes from using captions as lexical supervision for image-side sparse representations.

\begin{table}[h!]
\centering
\small
\setlength{\tabcolsep}{5pt}
\begin{tabular}{lcc}
\toprule
\textbf{Caption generator} & \textbf{NDCG@5} & \textbf{\flops{}} \\
\midrule
Li-LSR only           & .549 & 3.06 \\
\midrule
Qwen3-VL-2B           & .572 & 1.15 \\
Qwen3-VL-4B           & .568 & 1.01 \\
Qwen3-VL-8B           & .565 & 1.00 \\
Qwen3-VL-30B          & .564 & 1.10 \\
Qwen3-VL-235B         & .562 & 0.97 \\
\bottomrule
\end{tabular}
\caption{Caption-generator scale vs.\ retrieval quality (6-benchmark avg).}
\label{tab:caption_scale}
\end{table}

In summary, caption-gated token supervision improves lexical grounding by reinforcing retrieval-relevant image-side vocabulary dimensions, outperforming alternative supervision strategies and remaining robust across caption generators.

% Force all experiment-section floats to land before the Conclusion so
% tables don't drift onto an otherwise-empty page beyond the conclusion.
\FloatBarrier

\section{Conclusion}
\label{sec:conclusion}

Visual document retrieval still lacks a deployable lexical retriever for production-scale search.
Multimodal learned sparse retrieval naturally fits this missing operating point, but suffers from a lexical grounding challenge: visual pages do not explicitly indicate which vocabulary dimensions should be activated.
We address this challenge with \method{}, which uses caption-gated token supervision to guide image-side vocabulary activation during training.

Across standard benchmarks and an 18.7M-document corpus, \method{} outperforms the main baselines in its target regime without neural query encoding, and builds indexes substantially faster than OCR- or caption-based lexical pipelines.
It also complements neural retrievers through score fusion and as a first-stage retriever for multi-vector reranking.
Taken together, these results position \method{} as a lexical retrieval layer for the missing deployment setting in visual-document retrieval: query-encoding-free serving over directly indexed visual documents, with strong standalone performance and complementarity to dense and multi-vector systems.

\paragraph{Limitations and Future Work.}
Our study has four main limitations that we view as natural directions for future work.
First, we restrict the evaluation to English-language visual-documents; how the same caption-gated supervision behaves under multilingual retrieval remains to be studied.
Second, we deliberately focus on a compact, efficient sub-billion-parameter sparse model, leaving open how the approach scales when applied on top of substantially larger backbones.
Third, our analysis is limited to visual-document retrieval; whether the same lexical grounding mechanism transfers to broader multimodal tasks (e.g., natural-image retrieval or video) is an open question.
Fourth, we use a fixed captioning setup; studying prompt-controlled captions that target specific document regions, structures, or domain cues is a promising direction for specialized visual-document tasks.

\clearpage

\section*{GenAI Usage Disclosure}
Generative AI tools were used to improve the clarity and style of the writing and to assist with code drafting and debugging. All technical content, experiments, code, and results were reviewed and validated by the authors.

\balance
\bibliographystyle{ACM-Reference-Format}
\bibliography{custom}

\appendix
% Appendix content moved to main body sections

\end{document}